# Long-distance near-field energy transport via propagating surface waves

Jian Dong[a], Junming Zhao[a, *], Linhua Liu[a, b, †]

[a] *School of Energy Science and Engineering, Harbin Institute of Technology, Harbin 150001, China*
[b] *Department of Physics, Harbin Institute of Technology, Harbin 150001, China*

**Abstract**

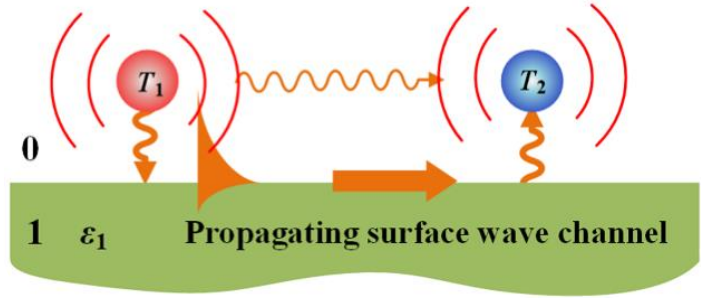

Near-field radiative heat transfer (RHT) between two bodies can significantly exceed the far-field limit set by Planck's law due to the evanescent wave tunneling, which typically can only occur when the two bodies are separated at subwavelength distances. We show that the RHT between two SiC nanoparticles with separation distances much larger than the thermal wavelength can still exceed the far-field limit when the particles are located within a subwavelength distance away from a SiC substrate. In this configuration, the localized surface phonon polariton (SPhP) of the particles couples to the propagating SPhP of the substrate which then provides a new channel for the near-field energy transport and enhances the RHT by orders of magnitude at large distances. The enhancement is also demonstrated to appear in a chain of closely spaced SiC nanoparticles located in the near field of a SiC substrate. The findings provide a new way for the long-distance transport of near-field energy.

## I. Introduction

Since the prediction by Polder and Van Hove[1], it is now well established that near-field radiative heat transfer (RHT) can significantly exceed the far-field limit set by Planck's blackbody law due to the contribution of evanescent waves. Near-field RHT typically occurs only at subwavelength distances. Despite this restriction, many experiments regarding near-field RHT have been successfully performed [2-

---

Authors to whom correspondence should be addressed:
* jmzhao@hit.edu.cn (Junming Zhao),
† lhliu@hit.edu.cn (Linhua Liu)

8], bringing more hope to the practical applications of near-field RHT like near-field energy conversion [9], thermal rectification [10], thermal logic [11], etc. Near-field RHT is still a vital field of research. Recent theoretical studies showed that the near-field RHT between two objects can be tuned and enhanced by tailoring the surface structures [12-15]. Near-field RHT can also be enhanced through many-body interactions [16-19]. The heat flux between two slabs can be amplified by an intermediate slab having coupled resonances with the two slabs [16]. Under similar conditions, the near-field RHT between two particles can be enhanced by several times when a third particle is placed between them [17-19].

An important question for near-field RHT is if the enhancement can be realized at larger separation distances than the characteristic thermal wavelength. Recently, several methods have been proposed that enables the long-distance transport of near-field energy. Nefedov and Simovski [20] demonstrated that the RHT across a micro-gap thermophotovoltaic system can be increased by orders of magnitude when the gap is filled with hyperbolic material that converts evanescent waves (waves with wave vectors larger than $\omega/c$) into propagating ones. Messina et al. [21] proposed that the near-field energy can be transported to distances much larger than the thermal wavelength when the two bodies exchanging thermal energy are connected in the near field by a weakly dissipating hyperbolic waveguide. Besides, Müller et al. [22] found that the RHT between two plates separated by distances larger than the thermal wavelength can be enhanced by orders of magnitude if the vacuum gap is filled with a nonabsorbing background medium that converts part of the evanescent waves into propagating ones. Using scattering theory, Asheichyk et al. [23] showed that the RHT between two point particles can be enhanced by orders of magnitude at large separation distances when the two particles are connected in the near field by a large sphere.

In this work, we present a new way for the long-distance transport of near-field energy. The near-field RHT between two particles is dominated by the strong interparticle coupling [Fig. 1(a)], where both propagating and evanescent waves contribute to the RHT. For the far-field RHT between the particles, only propagating waves contribute to the RHT [Fig. 1(b)]. We will demonstrate, as illustrated in Fig. 1(c), that two particles located within a subwavelength distance away from a substrate can exhibit long-distance exchange of near-field energy if propagating surface waves are excited. The propagating surface waves provide a new channel for the near-field energy transport. In addition to the configuration considered in this work, propagating surface waves can be supported by a wide range of materials and guided by miscellaneous means like thin film [24], stripe [25], nanowire [26] and graphene nanoribbon [27], which shows prospect for the long-distance transport of near-field energy.

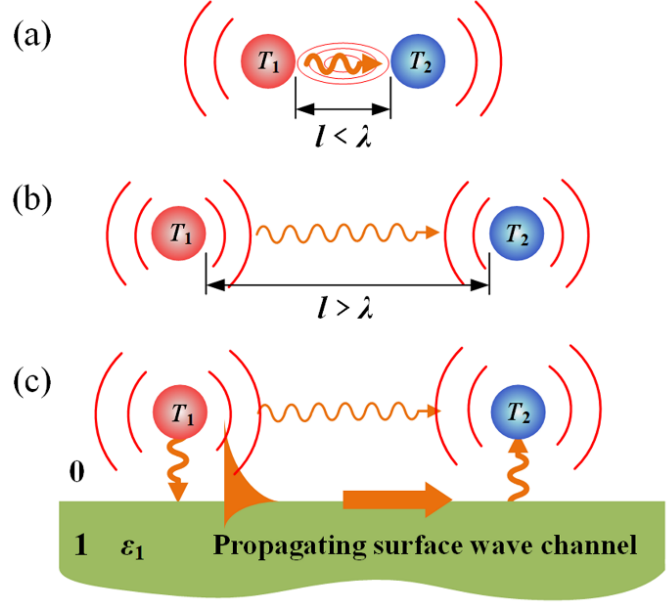

**Fig. 1** Radiative heat transfer (RHT) between two particles, (a) near-field RHT with strong interparticle coupling; (b) far-field radiative RHT; (c) when the particles are located in the near field of a substrate, propagating surface waves can provide a new channel for near-field energy transport.

## II. Theoretical aspects

To start the analysis, we first consider the many-body RHT involving a substrate as shown in Fig. 1(c). A general framework regarding the RHT between arbitrarily shaped objects and a surface has been proposed by Edalatpour and Francoeur [28], in which the surface interactions are considered via Sommerfeld's theory of electric dipole radiation above an infinite plane [29] and the volume integral equation of the electric field is solved using the thermal discrete dipole approximation [30]. Here we consider a relatively simple case. We suppose the particles are isotropic, linear, nonmagnetic, and the sizes of the particles are much smaller than the thermal wavelength $\lambda_T = c\hbar/k_B T$. In addition, we suppose that the separation distances from particle to particle and from particle to the substrate are sufficiently large so that higher multipoles can be neglected. Upon such simplifications, the particles can be modeled by point-like dipoles.

Assuming that the upper half space is vacuum, the total dipole moment of the *i*-th particle can be decomposed into the self-fluctuating part and the induced part

$$\mathbf{p}_i = \mathbf{p}_i^{fluc} + \varepsilon_0 \alpha_i \mathbf{E}_i^{inc} \tag{1}$$

where $\varepsilon_0$ is the vacuum permittivity and $\alpha_i$ is the electric polarizability of the *i*-th particle. For isotropic spherical nanoparticles, the electric polarizability can be obtained from the extinction cross section in the Mie theory [31,32]

$$\alpha_i = i \frac{6\pi}{k_0^3} a_1 \tag{2}$$

where $a_1$ is the first order of the Mie coefficient (see Appendix A). Note that the radiation damping of the particle is included in the electric polarizability. Assuming no external incident field, $\mathbf{E}_i^{inc}$ is the sum of the radiated and reflected fields from the particles and the radiated fields from the substrate

$$\mathbf{E}_i^{inc} = \omega^2 \mu_0 \sum_{j \neq i} \mathbf{G}_{ij}^0 \mathbf{p}_j + \omega^2 \mu_0 \sum_j \mathbf{G}_{ij}^R \mathbf{p}_j + \mathbf{E}_i^{sub} \tag{3}$$

$\mathbf{G}_{ij}^0 = \mathbf{G}^0(\mathbf{r}_i, \mathbf{r}_j)$ denotes the Green's tensor in vacuum

$$\mathbf{G}^0(\mathbf{r}_i, \mathbf{r}_j) = \frac{\exp(ik_0 r_{ij})}{4\pi r_{ij}} \left[ \left(1 + \frac{ik_0 r_{ij} - 1}{k_0^2 r_{ij}^2}\right) \mathbb{I}_3 + \frac{3 - 3ik_0 r_{ij} - k_0^2 r_{ij}^2}{k_0^2 r_{ij}^2} \hat{\mathbf{r}}_{ij} \otimes \hat{\mathbf{r}}_{ij} \right] \tag{4}$$

in which $k_0$ is the wave vector in vacuum, $r_{ij} = |\mathbf{r}_{ij}|$ is the magnitude of the vector linking $\mathbf{r}_i$ and $\mathbf{r}_j$, and $\hat{\mathbf{r}}_{ij} = \mathbf{r}_{ij}/r_{ij}$. $\mathbf{G}_{ij}^R = \mathbf{G}^R(\mathbf{r}_i, \mathbf{r}_j)$ is the reflection Green's tensor relating the electric field at $\mathbf{r}_i$ generated from the source at $\mathbf{r}_j$ through the surface reflection, which is given by the Sommerfeld integrals [33,34]

$$\mathbf{G}^R(\mathbf{r}_i, \mathbf{r}_j) = \int_0^\infty \frac{k_\rho dk_\rho}{4\pi \eta_0} \exp\left[-\eta_0 (z_i + z_j)\right] \mathbf{S}^{-1} \mathbf{F}(k_\rho, \rho_{ij}) \mathbf{S} \tag{5}$$

where $\rho_{ij}$ is the magnitude of the vector $\boldsymbol{\rho}_{ij} = (x_i - x_j)\hat{x} + (y_i - y_j)\hat{y}$, $x, y, z$ are the Cartesian components of the position vector $\mathbf{r}$, $\eta_0 = \sqrt{k_\rho^2 - k_0^2}$ and $k_\rho$ is the wave vector component parallel to the surface. $\mathbf{S}$ is the Jacobi rotation matrix given by [33,34]

$$\mathbf{S} = \frac{1}{k_\rho} \begin{bmatrix} k_x & k_y & 0 \\ -k_y & k_x & 0 \\ 0 & 0 & k_\rho \end{bmatrix} \tag{6}$$

where $k_{a=x,y}$ are the corresponding Cartesian components of the wave vector. The 3×3 tensor $\mathbf{F}(k_\rho, \rho_{ij})$ has 5 nonzero components, namely $F_{xx}$, $F_{yy}$, $F_{zz}$, $F_{xz}$ and $F_{zx}$. The $zz$ component, for instance, is given by [34]

$$F_{zz} = -\frac{\eta_1 - \varepsilon_1 \eta_0}{\eta_1 + \varepsilon_1 \eta_0} \frac{k_\rho^2}{k_0^2} J_0(k_\rho \rho_{ij}) \tag{7}$$

where $\varepsilon_1$ is the relative permittivity of the substrate, $\eta_1 = \sqrt{k_\rho^2 - k_1^2}$, and $J_0$ is the Bessel function of the first kind. A full description of the reflection Green's tensor can be found in Appendix B. It is noted that the integrand in the Sommerfeld integrals [Eq. (5)] is rapidly oscillating. To obtain sufficiently accurate results, we implemented the adaptive Gaussian-Kronrod quadrature to evaluate the integrals on the basis of the codes provided in Ref. [34].

Eqs. (1) and (3) can be casted in a matrix form as

$$\mathbb{A}\begin{pmatrix}\mathbf{p}_1\\\vdots\\\mathbf{p}_N\end{pmatrix}=\begin{pmatrix}\mathbf{p}_1^{fluc}\\\vdots\\\mathbf{p}_N^{fluc}\end{pmatrix}+\varepsilon_0\mathbb{O}\begin{pmatrix}\mathbf{E}_1^{sub}\\\vdots\\\mathbf{E}_N^{sub}\end{pmatrix} \tag{8}$$

where the elements of the interaction matrix $\mathbb{A}$ is given by $\mathbb{A}_{ij}=\delta_{ij}\mathbb{I}_3-(1-\delta_{ij})k_0^2\alpha_i\mathbf{G}_{ij}^0-k_0^2\alpha_i\mathbf{G}_{ij}^R$, and the elements of the matrix $\mathbb{O}$ is given by $\mathbb{O}_{ij}=\delta_{ij}\alpha_i\mathbb{I}_3$. Let $\mathbb{M}=\mathbb{A}^{-1}$ and $\mathbb{N}=\varepsilon_0\mathbb{A}^{-1}\mathbb{O}$, one has

$$\mathbf{P}=\mathbb{M}\mathbf{P}^{fluc}+\mathbb{N}\mathbf{E}^{sub} \tag{9}$$

where $\mathbf{P}$, $\mathbf{P}^{fluc}$ and $\mathbf{E}^{sub}$ are the corresponding vectors in Eq. (8). Thus, the total dipole moment of particle $i$ can be re-written as

$$\mathbf{p}_i=\mathbb{M}_{ii}\mathbf{p}_i^{fluc}+\sum_{j\neq i}\mathbb{M}_{ij}\mathbf{p}_j^{fluc}+\sum_j\mathbb{N}_{ij}\mathbf{E}_j^{sub} \tag{10}$$

The first term on the right hand side of Eq. (10) is the self-fluctuating part, the second term is the induced part by the fluctuating incident field from other particles, and the third term is the induced part by the incident field from the substrate.

The power absorption of particle $i$ due to external incident field is calculated by [32,35]

$$P_i(\omega)=\frac{1}{2}\frac{\omega}{\varepsilon_0}\bar{\chi}_i\text{Tr}\langle\mathbf{p}_i^{ind}(\omega)\mathbf{p}_i^{ind*}(\omega')\rangle \tag{11}$$

where $\bar{\chi}_i=\text{Im}\left[\left(\alpha_i^{-1}\right)^*\right]-k_0^3/6\pi$ and $\mathbf{p}^{ind}$ denotes the induced dipole moment. Applying the fluctuation dissipation theorem (FDT) [36]

$$\langle\mathbf{p}_{j,\beta}^{fluc}(\omega)\mathbf{p}_{j',\beta'}^{fluc*}(\omega')\rangle=\frac{4\varepsilon_0}{\pi\omega}\chi_j\Theta(\omega,T_j)\delta_{jj'}\delta_{\beta\beta'}\delta(\omega-\omega') \tag{12}$$

the radiative power from particle $j$ to particle $i$ is given by

$$P_{ij}=\frac{2}{\pi}\int_0^\infty d\omega\bar{\chi}_i\chi_j\text{Tr}\left[\mathbb{M}_{ij}\mathbb{M}_{ij}^*\right]\Theta(\omega,T_j) \tag{13}$$

where $\chi_j=\text{Im}(\alpha_j)-k_0^3|\alpha_j|^2/6\pi$ [36] and $\Theta(\omega,T)=\hbar\omega/(e^{\hbar\omega/k_BT}-1)$ is the mean energy of the Planck oscillator at the temperature $T$, $\hbar$ is the reduced Planck constant, $k_B$ is the Boltzmann's constant. It is noted that Eq. (13) is similar to that obtained in Refs. [17] and [18] in the absence of the substrate. For a point $\mathbf{r}_0$ above the surface and outside the particles, the electric field generated from the particles is

$$\mathbf{E}_{r_0}=\sum_j\omega^2\mu_0\left(\mathbf{G}_{r_0j}^0+\mathbf{G}_{r_0j}^R\right)\mathbb{M}\mathbf{p}_j^{fluc} \tag{14}$$

Defining $\left(\mathbb{Q}_{r_0 1} \cdots \mathbb{Q}_{r_0 N}\right) = \omega^2 \mu_0 \left[\left(\mathbf{G}^0_{r_0 1} + \mathbf{G}^R_{r_0 1}\right) \cdots \left(\mathbf{G}^0_{r_0 N} + \mathbf{G}^R_{r_0 N}\right)\right] \mathbb{M}$ and considering the FDT, the electric energy density emitted by the particles in the upper half space is expressed as

$$u_e(\mathbf{r}_0, \omega) = \frac{\varepsilon_0}{2} \langle \mathbf{E}_{r_0} \cdot \mathbf{E}^*_{r_0} \rangle = \frac{2\varepsilon_0^2}{\pi \omega} \sum_j \chi_j \Theta(\omega, T_j) \mathrm{Tr}\left[\mathbb{Q}_{r_0 j} \mathbb{Q}^*_{r_0 j}\right] \quad (15)$$

Note that the RHT between the substrate and the particles can also be derived following the procedure described in Ref. [28].

The RHT between two blackbody spheres is also considered for comparison. The total radiative heat flux between two blackbody spheres is calculated in the framework of the traditional radiation transfer theory

$$\mathcal{P}_{1-2} = \sigma\left(T_1^4 - T_2^4\right) A F_{1-2} \quad (16)$$

where $\sigma$ is the Stefan–Boltzmann constant, $T$ is the temperature, $A = 4\pi R^2$ and $R$ is the radius of the sphere. $F_{1-2}$ is the radiative view factor between the two spheres of equal radius, which is calculated from [37]

$$F_{1-2} = \frac{1}{2}\left\{1 - \left[1 - \frac{1}{(l/R+2)^2}\right]^{1/2}\right\} \quad (17)$$

where $l$ is the distance between two spheres edge to edge.

## III. Results and analysis

We first study the RHT between two SiC nanoparticles located near a substrate and show that the propagating surface waves can provide a new channel for near-field energy transport. Then, we will analyze the characteristics of the propagating surface wave channel for the RHT between two Ag nanoparticles and the RHT through a chain of closely spaced SiC nanoparticles.

SiC is a typical polar dielectric material, the dielectric function of which can be described by the Drude-Lorentz model

$$\varepsilon(\omega) = \varepsilon_\infty \frac{\omega^2 - \omega_l^2 + i\Gamma\omega}{\omega^2 - \omega_t^2 + i\Gamma\omega} \quad (18)$$

where $\varepsilon_\infty = 6.7$, $\Gamma = 0.9 \times 10^{12}$ rad·s$^{-1}$, and $\omega_t = 1.495 \times 10^{14}$ rad·s$^{-1}$ is the frequency of the transverse optical phonon, $\omega_l = 1.827 \times 10^{14}$ rad·s$^{-1}$ is the frequency of the longitudinal optical phonon [38]. Polar dielectrics can support SPhP in the region between $\omega_t$ and $\omega_l$, i.e., the so-called Reststrahlen band, where the permittivity becomes negative. The dielectric function of Ag can be described by the Drude model

$$\varepsilon(\omega) = 1 - \frac{\omega_p^2}{\omega^2 + i\Gamma\omega} \quad (19)$$

where $\omega_p = 1.37 \times 10^{16}$ rad·s$^{-1}$ is the plasmon frequency and $\Gamma = 2.73 \times 10^{13}$ rad·s$^{-1}$ [39]. The surface modes supported by Ag surface and nanoparticle lie in the ultraviolet range, which are far from typical thermal wavelength range.

## A. RHT between two SiC nanoparticles near a substrate

The RHT between two spherical SiC nanoparticles of radius $R=100$nm is studied in this sub-section. The validity of the dipole approximation for SiC nanoparticles with respect to the particle-particle and the particle-surface gaps has been examined by comparing to exact methods [40,41] in Appendix C. For particle-particle and particle-surface gaps larger than $3R=300$nm, the relative errors of the dipole approximation are within 10%. Thus, we maintain a minimum gap of 300nm from particle to particle and from particle to surface, which is enough to reproduce the general physics. Figure 2 shows the total heat flux between two SiC nanoparticles as a function of the separation distance edge to edge. The RHT between two SiC nanoparticles located in vacuum (none substrate), located above SiC and Ag substrates are considered. In addition, the far-field prediction of the heat flux between two blackbody spheres of equal size is considered for comparison.

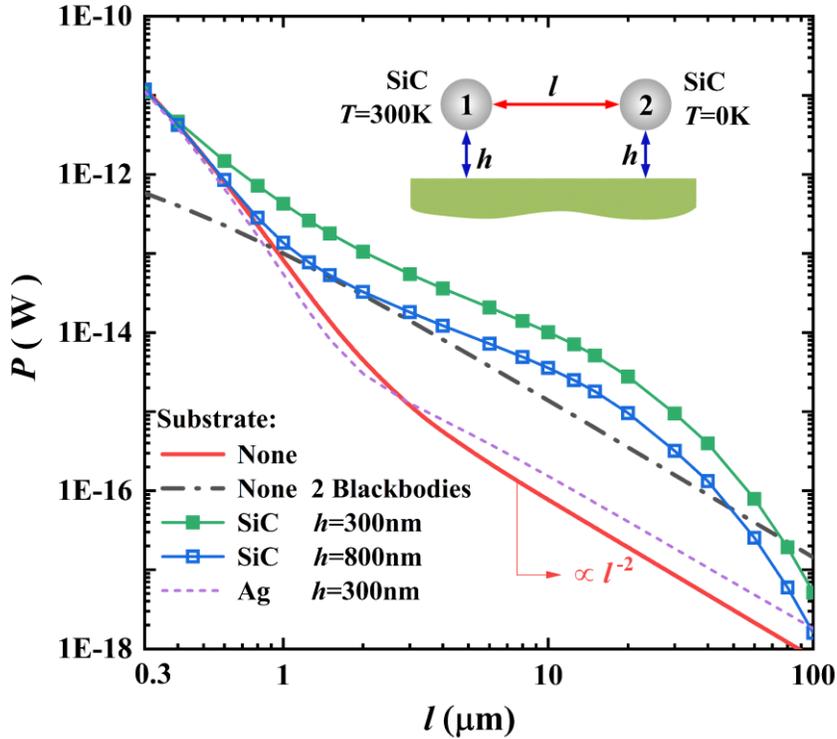

**Fig. 2** The total radiative heat flux $P$ (W) between two SiC nanoparticles of radius $R=100$nm as a function of the separation distance edge to edge $l$. Two SiC nanoparticles located in vacuum (none substrate), located above SiC and Ag substrates, and two equally sized blackbody spheres are considered. $h$ is the minimum gap between the particle and the substrate. The temperatures of particles 1 and 2 are kept at 300K and 0K, respectively.

As shown in Fig.2, the heat flux between two SiC nanoparticles in vacuum (none substrate) decays rapidly with increasing separation distance, and for separation distances larger than the characteristic thermal wavelength (about 7.63μm at 300K), the heat flux decays as $l^{-2}$, indicating that the RHT between the two particles reaches far-field mode [42]. When the two SiC nanoparticles are located at a distance $h$=300nm above a SiC substrate and separated by gaps smaller than 400nm, the heat flux shows no obvious differences from that in the absence of the substrate, which implies that the RHT is still dominated by the strong near-field interparticle coupling. With increasing separation distance between the two SiC nanoparticles, however, the heat flux is gradually enhanced by the SiC substrate. The enhancement of the heat flux gets larger with increasing *l*, and reaches a maximum at a separation distance of about 20μm. It can be seen that the heat flux is enhanced by more than two orders of magnitude for *l* ranging from 8 to 30μm. With further increase of the separation distance, however, the enhancement of the RHT decreases quickly. When the two SiC nanoparticles are located $h$=300nm above the SiC substrate, the heat flux between the two particles exceeds that predicted by the two blackbodies for *l* ranging from 300nm to as large as about 10 times the thermal wavelength. In the absence of the substrate, however, the heat flux between the two SiC nanoparticles can only exceed the far-field blackbody limit for separation distances smaller than 1μm. When the gap between SiC nanoparticle and the SiC substrate is increased to $h$=800nm, similar phenomena can be observed, but the enhancement in the RHT gets smaller. In this case, the heat flux between the two SiC nanoparticles can be enhanced by more than one order of magnitude for *l* ranging from several microns to about 60μm. For comparison, we also consider the RHT between two SiC nanoparticles located $h$=300nm above the Ag substrate that does not support surface waves in the thermal wavelength range. The Ag substrate has much smaller effects on the heat flux between the two SiC nanoparticles. It can be seen that the heat flux between the two SiC nanoparticles is decreased for *l* smaller than 2μm but is increased for larger separation distances.

To explain the phenomena observed in Fig. 2, we consider the spectral heat flux between two SiC nanoparticles. In the Reststrahlen band of SiC, the vacuum-SiC interface can support propagating surface waves called SPhP that produce a peak at the frequency near $1.79 \times 10^{14}$ rad·s$^{-1}$ [25]. And according to the Fröhlich condition [25], SiC nanoparticles in vacuum can support localized SPhP at the frequency near $1.756 \times 10^{14}$ rad·s$^{-1}$. Propagating surface waves have large wave vectors due to their bound nature, and the near-field radiation of the nanoparticles contains waves with large wave vectors. Therefore, propagating SPhP will be excited when the SiC nanoparticles are located in the near-field of the SiC substrate.

Figure 3 shows the spectral heat flux from $1.4 \times 10^{14}$ rad·s$^{-1}$ to $2.0 \times 10^{14}$ rad·s$^{-1}$ between two SiC nanoparticles that are located in vacuum (none substrate) and located $h$=300nm above SiC and Ag substrates. Separation distances of $l$=300nm, 1μm and 4μm between the two SiC nanoparticles are considered. In addition, the right axis of Fig. 3(b) gives the normalized propagation length $L/\lambda = 1/[\text{Im}(K)\lambda]$ of the propagating SPhP along a SiC-vacuum interface, where $K$ is obtained by the dispersion relation $K = k_0\sqrt{\varepsilon_1(\omega)/[\varepsilon_1(\omega)+1]}$ [43] and $\lambda$ is the corresponding wavelength.

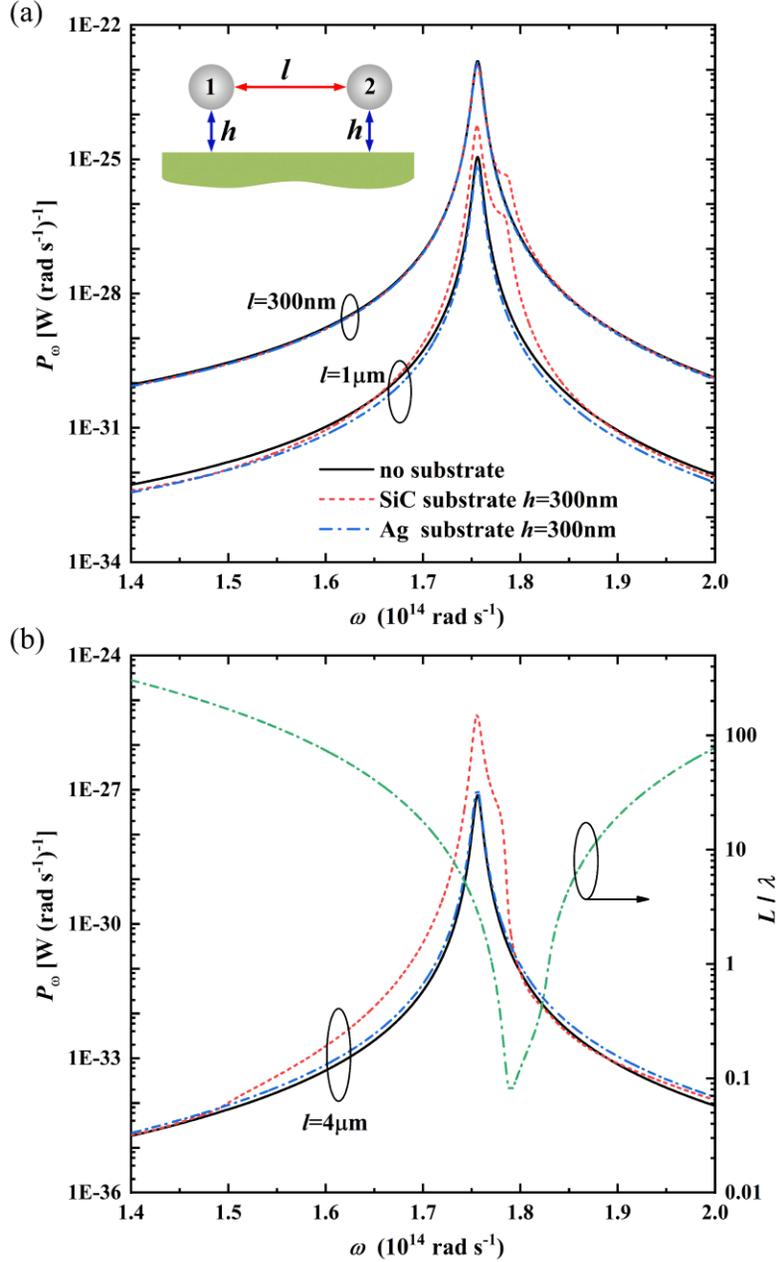

**Fig. 3** The spectral heat flux $P_\omega$ [W (rad s$^{-1}$)$^{-1}$] between two SiC nanoparticles of radius $R$=100nm that are located in vacuum (solid line) and located $h$=300nm above SiC (short dash) and Ag (dash dot) substrates. (a) $P_\omega$ for the particle-particle gaps of $l$=300nm and $l$=1μm; (b) $P_\omega$ for $l$=4μm, the normalized propagation length $L/\lambda$ [43] of the SPhP at the vacuum-SiC interface is plotted on the right axis. The temperatures of particles 1 and 2 are kept at 300K and 0K, respectively.

As shown in Fig. 3, the RHT between two SiC nanoparticles is dominated by the localized SPhP of the SiC nanoparticles. At other frequencies, however, the spectral heat flux can drop by orders of magnitude. When the two SiC nanoparticles separated by $l$=300nm are placed $h$=300nm above a SiC substrate, as shown in Fig. 3(a), the spectral heat flux between the two SiC nanoparticles is enhanced by several times near the frequency $1.79 \times 10^{14}$ rad·s$^{-1}$ at which the propagating SPhP of the SiC substrate produces a peak. This implies that the propagating SPhP on the SiC substrate is providing extra contribution to the RHT. However, the influences of the SiC substrate are not obvious near the frequency $1.756 \times 10^{14}$ rad·s$^{-1}$ since the RHT between the SiC nanoparticles is still dominated by the near-field interaction of the particles, i.e., the strong coupling of the localized SPhP. This explains why the RHT between two SiC nanoparticles is not obviously increased by the SiC substrate for $l$ smaller than 400nm as shown in Fig.2. When the separation distance increases to $l$=1μm, however, the SiC substrate enhances the spectral heat flux around $1.756 \times 10^{14}$ rad·s$^{-1}$ by several times. For $l$=4μm as shown in Fig. 3(b), the SiC substrate enhances the heat flux around $1.756 \times 10^{14}$ rad·s$^{-1}$ by more than one order of magnitude. The enhancement in the heat flux can be explained by the fact that the near-field coupling between the nanoparticles gets weaker quickly with increasing $l$, but in the presence of a SiC substrate the localized SPhP excites and couples with the propagating SPhP on the SiC substrate. The near-field energy of the SiC nanoparticles is transferred to that of the propagating SPhP that provides a new channel for the near-field energy transport. Yet the propagating SPhP decays along the SiC surface. As shown in Fig. 3(b), the propagating length $L$ of the SPhP on the vacuum-SiC interface is about twice the corresponding wavelength at $1.756 \times 10^{14}$ rad·s$^{-1}$ ($\lambda \approx 10.7$μm), which explains why the enhancement of the heat flux decreases quickly for separation distances larger than 20μm as shown in Fig. 2. When the SiC nanoparticles are located at $h$=300nm above an Ag substrate, the spectral heat flux between the SiC nanoparticles is not obviously influenced for $l$=300nm. For the separation distance $l$=1μm, the spectral heat flux is decreased by the Ag substrate, whereas for $l$=4μm the spectral heat flux is increased. To understand the effect of the Ag substrate on the RHT between SiC nanoparticles, more details are given in Appendix D.

Another example to show the coupling between the localized and propagating SPhP is the electric energy density distribution. Figure 4 shows the electric energy density distribution at the frequency $1.756 \times 10^{14}$ rad·s$^{-1}$ for two SiC nanoparticles located in vacuum and located at $h$=300nm above a SiC substrate, respectively. The separation distance between the two particles is $l$=4μm. The left particle is the emitter maintained at a temperature of 300K while the right particle is maintained at 0K. The emission of the SiC substrate is not considered. In the presence of the SiC substrate, a larger region of high energy density around the SiC nanoparticles can be observed, indicating the strong coupling between the localized SPhP and the propagating SPhP. In addition, due to the existence of the propagating SPhP on the substrate, the energy density along the surface is orders of magnitude larger than the same positions in the absence of the substrate.

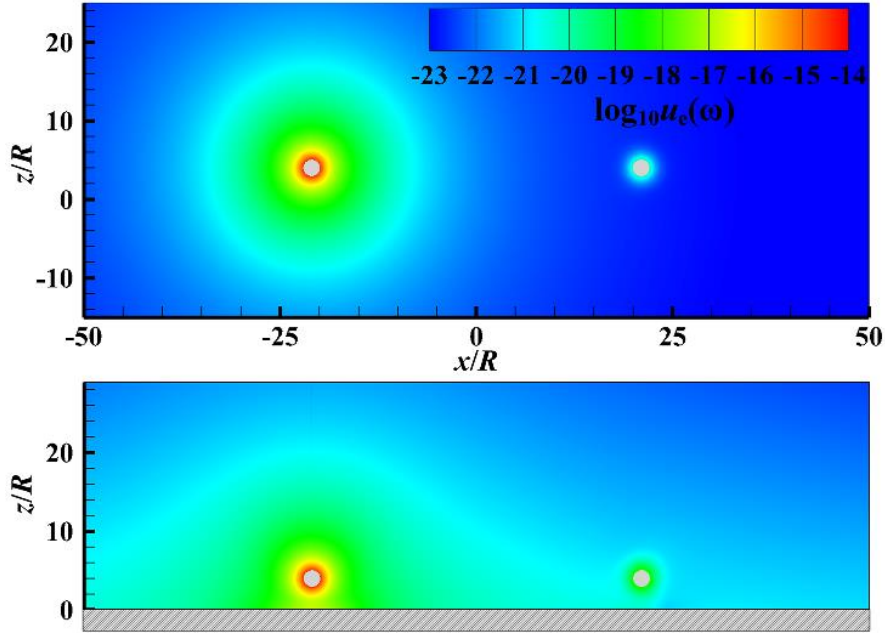

**Fig. 4** The electric energy density $u_e$ [J m$^{-3}$ (rad s$^{-1}$)$^{-1}$] distribution at $\omega = 1.756\times10^{14}$ rad·s$^{-1}$ for two SiC nanoparticles of radius $R$=100nm in vacuum (top) and located $h$=300nm above a SiC substrate (bottom). The gap between the two particles is $l$=4μm. The temperatures of the left and right particles are kept at 300K and 0K, respectively.

### B. The propagating surface wave channel for two Ag nanoparticles

We have shown that the localized SPhP of the SiC nanoparticles can excite and couple with the propagating SPhP of the SiC substrate when the particles are located within a subwavelength distance away from the substrate. Since the localized SPhP of the SiC nanoparticles makes the main contribution to the RHT, the propagating surface wave channel can enhance the total heat flux by orders of magnitude for large separation distances between the two SiC nanoparticles. One might wonder what the effect of the propagating surface wave channel will be if there is no coupling between localized and propagating surface modes. In this subsection, we consider the RHT between two Ag nanoparticles that are also located within a subwavelength distance from a SiC substrate. The surface mode of Ag nanoparticles in vacuum lies in the ultraviolet range, which is far from the Reststrahlen band of the SiC substrate.

Figure 5 shows the heat flux between two Ag nanoparticles in vacuum and located at $h$=300nm from a SiC substrate. The radius of the Ag nanoparticles is supposed to be 5nm, while the minimum particle-particle and particle-surface gaps are still maintained at 300nm. As shown in Fig. 5(a), the SiC substrate can enhance the total heat flux between two Ag nanoparticles for particle-particle gaps larger than 1μm, but the enhancement is much smaller than that observed for SiC nanoparticles. Figure 5(b) gives the spectral heat flux between two Ag nanoparticles separated by a distance of 2μm. As can be seen, the spectral heat flux can be enhanced by more than two orders of magnitude around the frequencies at which the SiC substrate supports propagating

SPhP. However, the spectral distribution of the heat flux between Ag nanoparticles is relatively flat in the thermal wavelength range, whereas the enhancement of the heat flux by the propagating SPhP lies in a very narrow spectral range. As a result, the enhancement of the total heat flux between Ag nanoparticles is much smaller than that observed between SiC nanoparticles.

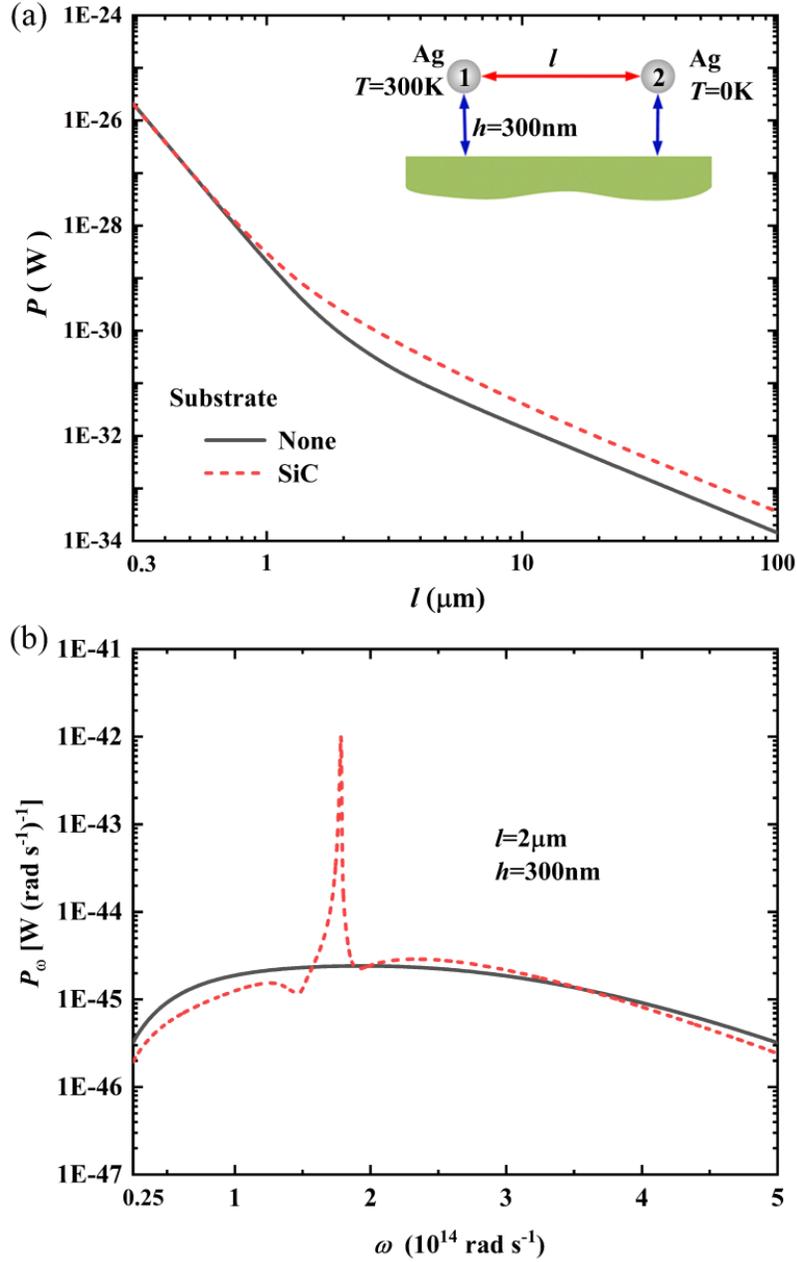

**Fig. 5** The heat flux between two spherical Ag nanoparticles of radius $R$=5nm in vacuum and located at a distance $h$=300nm from a SiC substrate, (a) the total heat flux $P$ (W) as a function of separation distance $l$; (b) the spectral heat flux $P_\omega$ [W (rad s$^{-1}$)$^{-1}$] for $l$=2μm. The temperatures of particles 1 and 2 are kept at 300K and 0K, respectively.

**C. The propagating surface wave channel for a chain of SiC nanoparticles**

For a chain of closely spaced nanoparticles, strong interparticle interactions can make the localized surface modes propagate along the chain, which has been studied widely for subdiffraction waveguiding [25,44] and electromagnetic energy transport [45,46]. Here, we show that the propagating surface wave channel can also enhance the RHT through a chain of closely spaced nanoparticles. Figure 6 shows the heat flux from particle 1 to 10 through a chain of 10 equidistantly distributed SiC nanoparticles with respect to the height above a SiC substrate. The radius of the SiC nanoparticles is 100nm, and the minimum particle-particle and particle-surface gaps are still kept at 300nm.

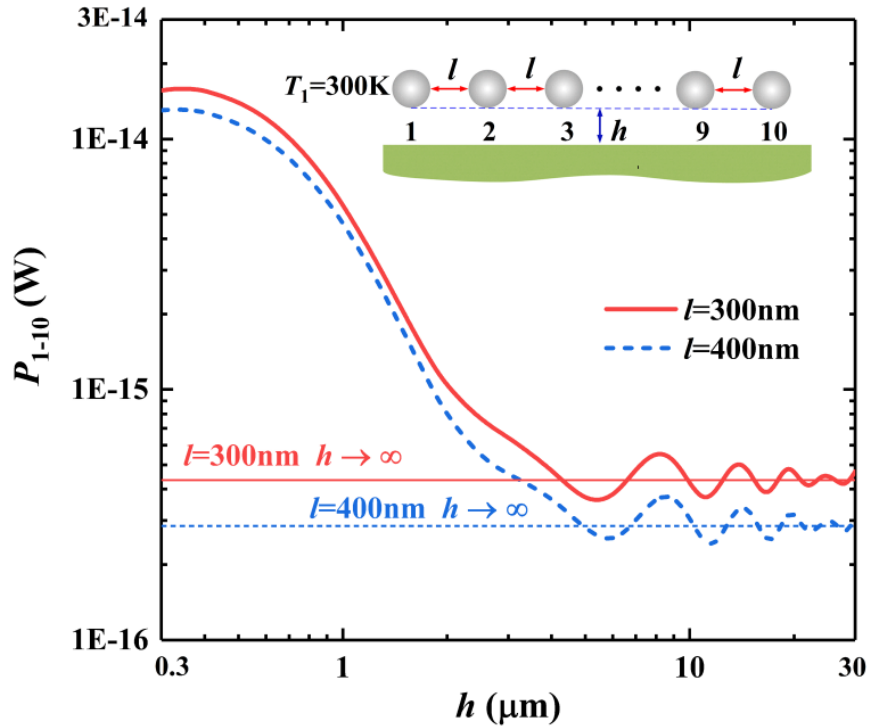

**Fig. 6** Radiative heat flux from particle 1 to particle 10 through a chain of 10 equidistantly distributed SiC particles as a function of the height $h$ above a SiC substrate. The temperature of particle 1 is kept at $T_1$=300K. The radius of all the nanoparticles is $R$=100nm.

As shown in Fig. 6, when the chain is located $h$=300nm above the SiC substrate, the heat flux from particle 1 to particle 10 is more than one order of magnitude larger than that without the substrate for interparticle distances of $l$=300nm and $l$=400nm. With increasing $h$, the heat flux decreases and drops by one order of magnitude when $h$ reaches 2μm, indicating that the propagating surface wave channel becomes weaker. For $h$ larger than 4μm, the heat flux shows damping oscillations with increasing $h$ due to the wave interference effect with the substrate, which can be understood by an inspection of the exponential term in the reflection Green's tensor. The heat flux gradually approaches the value in the absence of the substrate.

**IV Conclusion**

In conclusion, we have shown that the propagating surface waves can serve as a new channel for the transport of near-field energy at long distances. For two SiC nanoparticles with separation gaps larger than the thermal wavelength, the RHT can be enhanced by two orders of magnitude when the particles are located within subwavelength distances away from a SiC substrate. The giant enhancement in the RHT is attributed to the strong coupling of the localized SPhP of the particle to the propagating SPhP of the substrate which then provides a new channel for the long-distance transport of near-field energy. The propagating SPhP channel can also be observed for Ag nanoparticles that are modeled as electric dipoles. But the enhancement in the total heat flux is much smaller since there are no coupled surface modes between the Ag nanoparticle and the SiC substrate. In addition, the propagating SPhP channel can also enhance the heat flux through a chain of closely spaced SiC nanoparticles by one order of magnitude when they are located in the near field of a SiC substrate. The findings of this work provide a new way for the transport of near-field energy at long distances. It is noted that this work only considers spherical nanoparticles that are modeled as electric dipoles. The role of the propagating surface wave channel for particles with strong magnetic response, or for particles with more complex geometries or larger sizes still needs further study.

**Acknowledgments**

We acknowledge the support by National Natural Science Foundation of China (Nos. 51336002) and the Fundamental Research Funds for the Central Universities (Grant No. HIT.BRETIII.201415).

**Appendix A: The first order Mie coefficient**

The first order Mie coefficient is given by [31]

$$a_1 = \frac{\varepsilon j_1(y)\left[xj_1(x)\right]' - j_1(x)\left[yj_1(y)\right]'}{\varepsilon j_1(y)\left[xh_1^{(1)}(x)\right]' - h_1^{(1)}(x)\left[yj_1(y)\right]'} \quad (20)$$

where $x = kR$ and $y = \sqrt{\varepsilon}kR$, $R$ denotes the radius of the particle and $\varepsilon$ denotes the electric permittivity of the particle, $j_1$ is the first order Bessel function, and $h_1^{(1)}$ the first order Hankel function of the first kind [31].

**Appendix B: The reflection Green's tensor**

The reflection Green's tensor is given by [33,34]

$$\mathbf{G}^R(\mathbf{r}_i,\mathbf{r}_j) = \int_0^\infty \frac{k_\rho dk_\rho}{4\pi\eta_0} \exp\left[-\eta_0(z_i+z_j)\right] \mathbf{S}^{-1}\mathbf{F}(k_\rho,\rho_{ij})\mathbf{S} \quad (21)$$

where $\rho_{ij}$ is the magnitude of the vector $\boldsymbol{\rho}_{ij} = (x_i - x_j)\hat{x} + (y_i - y_j)\hat{y}$, $x$, $y$, $z$ are the Cartesian components of the position vector $\mathbf{r}$, $\eta_0 = \sqrt{k_\rho^2 - k_0^2}$, $k_\rho$ is the wave vector component parallel to surface and $k_0$ is the wave number in vacuum. $\mathbf{S}$ is the Jacobi rotation matrix which is given by [33]

$$\mathbf{S} = \frac{1}{k_\rho}\begin{bmatrix} k_x & k_y & 0 \\ -k_y & k_x & 0 \\ 0 & 0 & k_\rho \end{bmatrix} \quad (22)$$

$\mathbf{F}(k_\rho,\rho_{ij})$ is a 3×3 tensor which is given by [34]

$$\mathbf{F}(k_\rho,\rho_{ij}) = \begin{bmatrix} F_{xx} & 0 & F_{xz} \\ 0 & F_{yy} & 0 \\ F_{zx} & 0 & F_{zz} \end{bmatrix} \quad (23)$$

Defining [34]

$$\begin{aligned} d_{xx} &= -\frac{\eta_1 - \varepsilon_1\eta_0}{\eta_1 + \varepsilon_1\eta_0} \frac{\eta_0^2}{k_0^2} \\ d_{yy} &= \frac{\eta_0 - \eta_1}{\eta_0 + \eta_1} \\ d_{zz} &= -\frac{\eta_1 - \varepsilon_1\eta_0}{\eta_1 + \varepsilon_1\eta_0} \frac{k_\rho^2}{k_0^2} \\ d_{xz} &= -d_{zx} = \frac{\eta_1 - \varepsilon_1\eta_0}{\eta_1 + \varepsilon_1\eta_0} \frac{ik_\rho\eta_0}{k_0^2} \end{aligned} \quad (24)$$

where $\eta_1 = \sqrt{k_\rho^2 - k_1^2}$ and $k_1$ is the wave vector in the substrate, the components of $\mathbf{F}(k_\rho,\rho_{ij})$ are given by [34]

$$\begin{aligned} F_{xx} &= (d_{yy} - d_{xx})\frac{J_1(k_\rho\rho_{ij})}{k_\rho\rho_{ij}} + d_{xx}J_0(k_\rho\rho_{ij}) \\ F_{yy} &= (d_{xx} - d_{yy})\frac{J_1(k_\rho\rho_{ij})}{k_\rho\rho_{ij}} + d_{yy}J_0(k_\rho\rho_{ij}) \\ F_{zz} &= d_{zz}J_0(k_\rho\rho_{ij}) \\ F_{xz} &= -F_{zx} = id_{xz}J_1(k_\rho\rho_{ij}) \end{aligned} \quad (25)$$

where $J_0$ and $J_1$ are the Bessel functions of the first kind.

**Appendix C: The dipole approximation of the SiC nanoparticle with respect to the particle-particle and particle-surface gaps**

In the dipole approximation of nanoparticles, it is required that the separation distances from particle to particle and from particle to surface are sufficiently large so that higher multipoles can be neglected. To verify the minimum particle-particle and particle-surface gaps considered in this study, we compared the dipole approximation of spherical SiC nanoparticles with exact solutions of the RHT between two spheres [40] and the RHT between a sphere and a substrate [41]. Using the formulas in this work, the dipole approximation for the RHT between two nanoparticles can be easily calculated without considering the effect of the substrate. And following the procedure in Ref. [28], the dipole approximation for the radiative heat flux between a nanoparticle and a substrate can be expressed as

$$\mathcal{P}_{i,sub} = \frac{1}{2\varepsilon_0} \int_0^\infty d\omega\, \omega \bar{\chi}_i \mathrm{Tr}\left[ \sum_{jj'} \mathbb{N}_{ij} \left\langle \mathbf{E}_j^{sub}(\omega) \mathbf{E}_{j'}^{sub*}(\omega') \right\rangle \mathbb{N}_{ij'}^* \right] \quad (26)$$

where $\mathbb{N}_{ij}$ is defined in the main text, and the expression for the spatial correlation function $\left\langle \mathbf{E}_j^{sub}(\omega) \mathbf{E}_{j'}^{sub*}(\omega') \right\rangle$ can be found in Ref. [28].

Figure 7 compares the dipole approximation and the exact solutions for the thermal conductance at 300K with respect to the particle-particle and particle-substrate gaps. The radius of the SiC nanoparticle is $R$=100nm. The thermal conductance is defined as [40]

$$G = \lim_{\Delta T \to 0} (\Delta \mathcal{P})/\Delta T \quad (27)$$

As shown in Fig. 7, for smaller particle-particle and particle-substrate gaps, the dipole approximation tends to underestimate the thermal conductance between two SiC nanoparticles and that between a SiC nanoparticle and a SiC substrate, since the contributions of higher multipoles are not included. With increasing separation gaps, the accuracy of dipole approximation gradually gets better. The relative errors of the dipole approximation are within 10% for particle-particle and particle-surface gaps larger than $3R$=300nm. Thus, for the cases considered in this work, it is enough to reveal the general physics to maintain a minimum particle-particle gap of $l$=$3R$=300nm and a minimum particle-substrate gap of $h$=$3R$=300nm. As to the Ag nanoparticles considered in this work, the radius is supposed to be 5nm while the minimum particle-particle and particle-substrate gaps are still kept at 300nm, the contribution of higher multipoles can also be neglected.

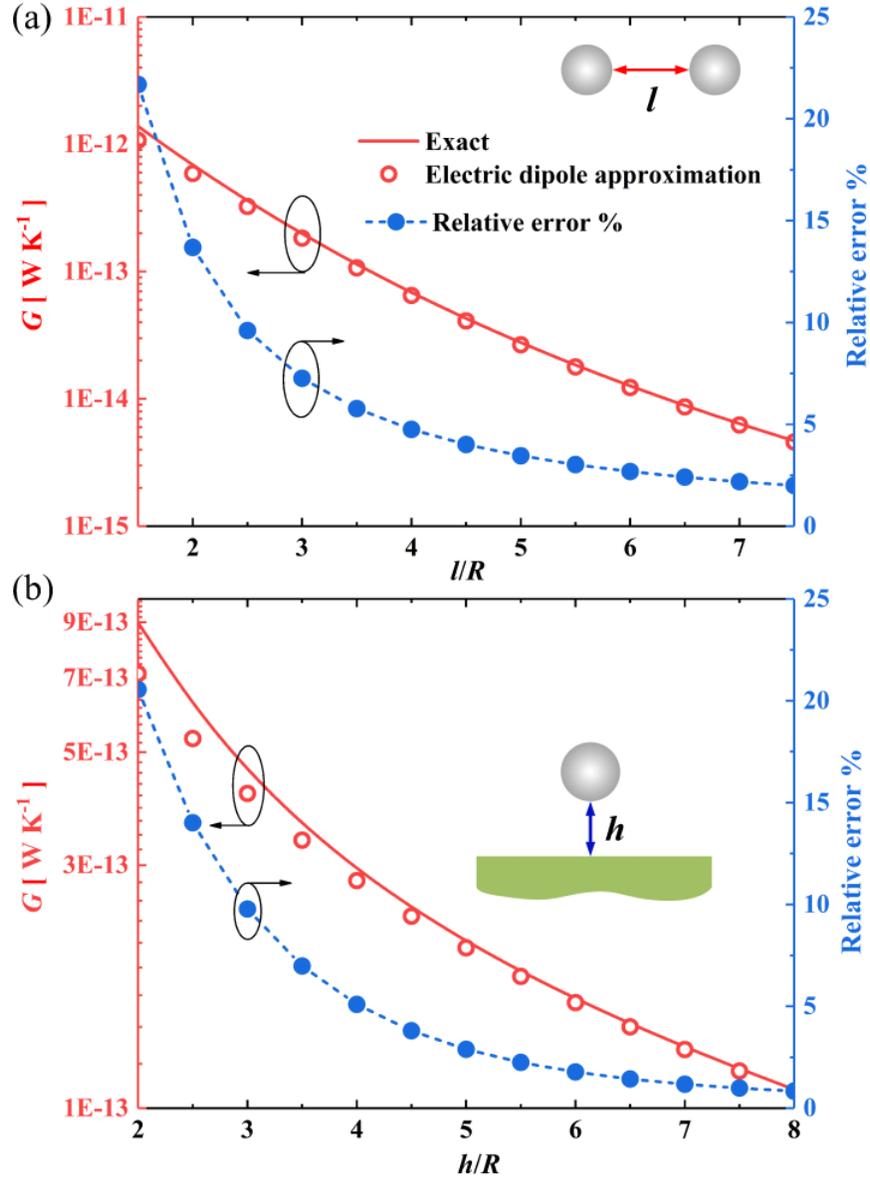

**Fig. 7** Comparison of the thermal conductance $G$ at 300K between the dipole approximation and the exact solutions, (a) $G$ between two spherical SiC nanoparticles of radius $R$=100nm as a function of the particle-particle gap; (b) $G$ between a spherical SiC nanoparticle of radius $R$=100nm and a SiC substrate as a function of the particle-substrate gap. The relative errors of the dipole approximation are plotted on the right axis.

**Appendix D Interactions between two SiC nanoparticles near the Ag substrate**

To understand the RHT between two SiC nanoparticles near the Ag substrate as shown in Figs. 2 and 3, we investigate the electromagnetic interactions between the two SiC nanoparticles near the Ag substrate. The fluctuating dipole moment of particle 1 given by the FDT is randomly oriented. For ease of analysis, we assign *x*, *y* and *z* polarized dipole moments to particle 1, respectively, and study how particle 2 responses to particle 1. To quantify the interactions, the transmission efficiency from particle 1 to particle 2 is applied, which is defined as [44]

$$\mathcal{T}_{21} = \left|\mathbf{p}_2^{ind}\right|^2 / \left|\mathbf{p}_1\right|^2 \tag{28}$$

For a given dipole moment $\mathbf{p}_1$ of particle 1, the induced dipole moment in particle 2 can be calculated according to Eq. (10) in the main text as follows

$$\mathbf{p}_2^{ind} = \mathbb{M}_{21}\mathbf{p}_1 \tag{29}$$

Figure 8 (a) shows the transmission efficiency from particle 1 to particle 2 as a function separation distance *l* between the two SiC nanoparticles. In the absence of the substrate, the transmission efficiency is dominated by the *x* polarized dipole moment of particle 1 for *l* smaller than 1μm, but is dominated by the *y* and *z* polarized dipole moments for *l* larger than 4μm. This implies that the near-field RHT from particle 1 to particle 2 is dominated by the longitudinal component of the fluctuating dipole moment, while at larger separation distances the RHT is dominated by the transversal component of the fluctuating dipole moment. When the SiC nanoparticles are located near the Ag substrate, however, the transmission efficiency is decreased for *x* and *y* polarized dipole moment of particle 1, whereas it increases for *z* polarized dipole moment for *l* larger than 800nm. Actually, the Ag substrate behaves like a mirror due to its large permittivity. As illustrated in Fig. 8 (b), the dipole moments parallel (*x* and *y*) to the substrate will be weakened by the induced image dipole moments that have opposite orientations. However, the *z* polarized dipole moment will form an enlarged effective dipole moment with its image dipole moment having the same orientation. For the RHT between two SiC nanoparticles, therefore, the Ag substrate will decrease the contribution of the *x* and *y* components of the fluctuating dipole moment, whereas it can increase the contribution of the *z* component. Such a complex mechanism leads to the phenomena observed in Figs. 2 and 3, i.e., the Ag substrate can either decrease or increase the RHT between two SiC nanoparticles depending on the particle-particle gaps.

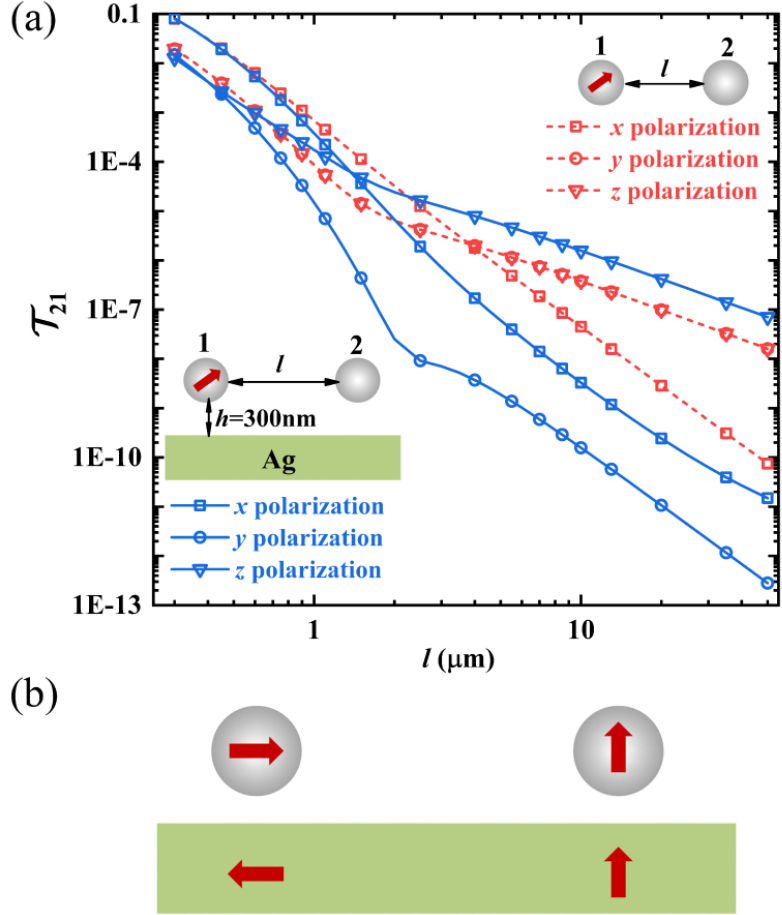

**Fig. 8** (a) The transmission efficiency $\mathcal{T}_{21}$ [see Eq.(28)] as a function of the separation distance *l* for *x*, *y* and *z* polarized dipole moment of particle 1. Two SiC nanoparticles located in vacuum (no substrate) and located *h*=300nm above the Ag substrate are considered; the radius of the particles is *R*=100nm and the frequency is $\omega = 1.756\times10^{14}$ rad·s$^{-1}$. (b) Schematic of the image dipole orientation for dipole moment in parallel and vertical orientations.

**References**

[1] D. Polder and M. Van Hove, Phys. Rev. B **4**, 3303 (1971).
[2] E. Rousseau, A. Siria, G. Jourdan, S. Volz, F. Comin, J. Chevrier, and J.-J. Greffet, Nat. Photonics **3**, 514 (2009).
[3] S. Shen, A. Narayanaswamy, and G. Chen, Nano Lett. **9**, 2909 (2009).
[4] R. St-Gelais, B. Guha, L. Zhu, S. Fan, and M. Lipson, Nano Lett. **14**, 6971 (2014).
[5] B. Song, Y. Ganjeh, S. Sadat, D. Thompson, A. Fiorino, V. Fernández-Hurtado, J. Feist, F. J. Garcia-Vidal, J. C. Cuevas, and P. Reddy, Nat. Nanotechnol. **10**, 253 (2015).
[6] M. P. Bernardi, D. Milovich, and M. Francoeur, Nat. Commun. **7**, 12900 (2016).

*Phys. Rev. B 2018. In press*[7] J. I. Watjen, B. Zhao, and Z. M. Zhang, Appl. Phys. Lett. **109**, 203112 (2016).

[8] K. Ito, K. Nishikawa, A. Miura, H. Toshiyoshi, and H. Iizuka, Nano Lett. (2017).

[9] A. Narayanaswamy and G. Chen, Appl. Phys. Lett. **82**, 3544 (2003).

[10] C. R. Otey, W. T. Lau, and S. Fan, Phys. Rev. Lett. **104**, 154301 (2010).

[11] P. Ben-Abdallah and S.-A. Biehs, Phys. Rev. Lett. **112**, 044301 (2014).

[12] R. Guérout, J. Lussange, F. S. S. Rosa, J.-P. Hugonin, D. A. R. Dalvit, J.-J. Greffet, A. Lambrecht, and S. Reynaud, Phys. Rev. B **85**, 180301(R) (2012).

[13] X. Liu and Z. Zhang, ACS Photonics **2**, 1320 (2015).

[14] J. Dai, F. Ding, S. I. Bozhevolnyi, and M. Yan, Phys. Rev. B **95**, 245405 (2017).

[15] V. Fernández-Hurtado, F. J. Garcia-Vidal, S. Fan, and J. C. Cuevas, Phys. Rev. Lett. **118**, 203901 (2017).

[16] R. Messina, M. Antezza, and P. Ben-Abdallah, Phys. Rev. Lett. **109**, 244302 (2012).

[17] P. Ben-Abdallah, S.-A. Biehs, and K. Joulain, Phys. Rev. Lett. **107**, 114301 (2011).

[18] J. Dong, J. Zhao, and L. Liu, Phys. Rev. B **95**, 125411 (2017).

[19] M. Nikbakht, J. Appl. Phys. **116**, 094307 (2014).

[20] I. S. Nefedov and C. R. Simovski, Phys. Rev. B **84**, 195459 (2011).

[21] R. Messina, P. Ben-Abdallah, B. Guizal, M. Antezza, and S.-A. Biehs, Phys. Rev. B **94**, 104301 (2016).

[22] B. Müller, R. Incardone, M. Antezza, T. Emig, and M. Krüger, Physical Review B **95**, 085413 (2017).

[23] K. Asheichyk, B. Müller, and M. Krüger, Phys. Rev. B **96**, 155402 (2017).

[24] S. Gluchko, B. Palpant, S. Volz, R. Braive, and T. Antoni, Appl. Phys. Lett. **110**, 263108 (2017).

[25] S. A. Maier, *Plasmonics: fundamentals and applications* (Springer, New York, 2007).

[26] R. F. Oulton, V. J. Sorger, D. A. Genov, D. F. P. Pile, and X. Zhang, Nat. Photonics **2**, 496 (2008).

[27] J. Christensen, A. Manjavacas, S. Thongrattanasiri, F. H. L. Koppens, and F. J. García de Abajo, ACS Nano **6**, 431 (2011).

[28] S. Edalatpour and M. Francoeur, Phys. Rev. B **94**, 045406 (2016).

[29] A. Sommerfeld, Ann. Phys. **28**, 665 (1909).

[30] S. Edalatpour and M. Francoeur, J. Quant. Spectrosc. Radiat. Transf. **133**, 364 (2013).

[31] C. F. Bohren and D. R. Huffman, *Absorption and scattering of light by small particles* (John Wiley & Sons, New York, 1983).

[32] G. W. Mulholland, C. F. Bohren, and K. A. Fuller, Langmuir **10**, 2533 (1994).

[33] A. A. Maradudin and D. L. Mills, Phys. Rev. B **11**, 1392 (1975).

[34] G. Y. Panasyuk, J. C. Schotland, and V. A. Markel, J. Phys. A-Math. Theor. **42**, 275203 (2009).

[35] B. T. Draine, Astrophys. J. **333**, 848 (1988).

[36] R. Messina, M. Tschikin, S.-A. Biehs, and P. Ben-Abdallah, Phys. Rev. B **88**, 104307 (2013).

[37] J. R. Howell, R. Siegel, and M. P. Mengüç, *Thermal radiation heat transfer* (CRC Press, Boca Raton, 2015).

[38] E. D. Palik, *Handbook of Optical Constants of Solids* (Academic Press, New York, 1985).

[39] M. A. Ordal, L. L. Long, R. J. Bell, S. E. Bell, R. R. Bell, R. W. Alexander, and C. A. Ward, Appl. Optics **22**, 1099 (1983).

[40] A. Narayanaswamy and G. Chen, Phys. Rev. B **77**, 075125 (2008).
20

[41] C. Otey and S. Fan, Phys. Rev. B **84**, 245431 (2011).
[42] P.-O. Chapuis, M. Laroche, S. Volz, and J.-J. Greffet, Appl. Phys. Lett. **92**, 201906 (2008).
[43] K. Joulain, J.-P. Mulet, F. Marquier, R. Carminati, and J.-J. Greffet, Surf. Sci. Rep. **57**, 59 (2005).
[44] P. J. Compaijen, V. A. Malyshev, and J. Knoester, Phys. Rev. B **87**, 205437 (2013).
[45] J. Ordonez-Miranda, L. Tranchant, S. Gluchko, and S. Volz, Phys. Rev. B **92**, 115409 (2015).
[46] E. Tervo, Z. Zhang, and B. Cola, Physical Review Materials **1**, 015201 (2017).